# Fast near-infrared photodetectors based on nontoxic and solution-processable AgBiS$_2$

*Yi-Teng Huang*, *Davide Nodari*, *Francesco Furlan*, *Youcheng Zhang*, *Marin Rusu*, *Linjie Dai*, *Zahra Andaji-Garmaroudi*, *Samuel D. Stranks*, *Henning Sirringhaus*, *Akshay Rao*, *Nicola Gasparini*,* and *Robert L. Z. Hoye**

Dr. Y.-T. Huang, Prof. R. L. Z. Hoye
Inorganic Chemistry Laboratory, Department of Chemistry, University of Oxford, South Parks Road, Oxford OX1 3QR, United Kingdom
Email: robert.hoye@chem.ox.ac.uk

Dr. Y.-T. Huang, Y. Zhang, Dr. L. Dai, Dr. Z. Andaji-Garmaroudi, Prof. S. D. Stanks, Prof. H. Sirringhaus, Prof. A. Rao
Cavendish Laboratory, University of Cambridge, JJ Thomson Ave, Cambridge CB3 0HE, United Kingdom

D. Nodari, F. Furlan, Dr. N. Gasparini
Department of Chemistry and Centre for Processable Electronics, Imperial College London, White City Campus, London W12 0BZ, UK
Email: n.gasparini@imperial.ac.uk

Dr. M. Rusu
Struktur und Dynamik von Energiematerialien, Helmholtz-Zentrum Berlin für Materialien und Energie, 14109 Berlin, Germany

Prof. S. D. Stranks
Department of Chemical Engineering and Biotechnology, University of Cambridge, Philippa Fawcett Drive, CB3 0AS Cambridge, UK

Dr. R. L. Z. Hoye
Department of Materials, Imperial College London, Exhibition Road, London SW7 2AZ, United Kingdom

**Abstract**

Solution-processable near-infrared (NIR) photodetectors are urgently needed for a wide range of next-generation electronics, including sensors, optical communications and bioimaging. However, there is currently a compromise between low toxicity and slow (<300 kHz cut-off frequency) organic materials versus faster detectors (>300 kHz cut-off frequency) based on compounds containing toxic lead or cadmium. Herein, we circumvent this trade-off by developing solution-processed $AgBiS_2$ photodetectors with high cut-off frequencies under both white light (>1 MHz) and NIR (approaching 500 kHz) illumination. These high cut-off frequencies are due to the short transit distances of charge-carriers in the $AgBiS_2$ photodetectors, which arise from the strong light absorption of these materials, such that film thicknesses well below 120 nm are adequate to absorb >65% of near-infrared to visible light. By finely controlling the thickness of the photoactive layer, we can modulate the charge-collection efficiency, achieve low dark current densities, and minimize the effects of ion migration to realize fast photodetectors that are stable in air. These outstanding characteristics enable real-time heartbeat sensors based on NIR $AgBiS_2$ photodetectors.

## 1. Introduction

Near-infrared photodetectors (NIR PDs) have been gaining increasing attention over the past decade, owing to the increasing needs in automotive vehicles, smart phones, augmented reality, machine vision, biometric monitoring, and imaging.[1–4] Many high throughput applications, such as optical communications[5] and computed axial tomography[6] also require NIR PDs with fast photo-response. The speed of PD photo-response is usually characterized by the cut-off frequency ($f_{-3dB}$), which is highly associated with the rise and fall times of PDs.

In the wavelength range between 850-1100 nm, which is crucial for LiDAR (light detection and ranging) technology[7] and biomedical imaging,[8,9] silicon-based PDs dominate the current market due to their low cost and easy integration. However, silicon-based PDs are usually bulky and require sophisticated manufacturing processes, making them less attractive as portable or wearable devices. On the other hand, organic materials,[10–12] lead-halide perovskites,[7,13] and chalcogenide quantum dots (QDs)[14–16] have demonstrated higher absorption coefficients in the NIR region than silicon, and their solution processability at low temperature enable them to be compatible with polymer substrates, which are ideal for flexible and lightweight electronics. Nevertheless, NIR PDs based on these materials also face various challenges. For example, the performance of NIR organic PDs is limited either by the low charge-carrier mobility or high non-radiative recombination losses,[17] while perovskite PDs are not stable in air and encapsulation is usually needed. Additionally, heavy metal contamination to the environment (*e.g.*, Pb or Cd pollution) remains a concern for PDs based on metal-halide perovskites (*e.g.*, Pb-Sn perovskites) and chalcogenide QDs (*e.g.*, PbS and CdTe QDs) used for NIR photodetection. Indeed, the usage of Pb and Cd in consumer electronics is regulated in many jurisdictions, such as by the EU Restriction of Hazardous Substances (RoHS) directive to up to 0.1 wt.% for Pb and 0.01 wt.% for Cd.[18] Apart from these challenges, another limiting factor is that very few of the above solution-processed PDs could achieve cut-off frequencies exceeding 300 kHz under visible light illumination, and it is even rarer in NIR region (Table S1, Supporting Information). So far, to our knowledge, the best example of a PD that has a cut-off frequency exceeding 300 kHz in the NIR region is based on toxic PbS QDs. The lack of options for fast, solution processable NIR PDs could be a great hindrance in high-speed applications, such as optical communications and data transfer.

AgBiS$_2$ is composed of RoHS-compliant elements with sufficient availability for commercial applications.[19] It has emerged as a promising absorber for photovoltaics, especially due to its high absorption coefficients exceeding those of standard inorganic thin film materials.[20–22] As a result, efficient photovoltaics are achievable using only 30 nm thick absorber layers.[21] Surprisingly, although the bandgap of AgBiS$_2$ (~1.0–1.3 eV) is also suitable for NIR light detection, there are only few investigations into AgBiS$_2$ NIR PDs. Some early studies have demonstrated that a high cut-off frequency in the visible region (630 nm wavelength) and a high specific detectivity $D^*$ in the NIR region could be achieved in PDs based on solution-processed bulk thin films of AgBiS$_2$[23] and AgBiS$_2$ nanocrystals (NCs),[24,25] respectively (Table S1, Supporting Information). Nevertheless, AgBiS$_2$ PDs that simultaneously exhibit outstanding photodetection along with a fast response in the NIR range have not yet been realized.

With escalating demand for ultrafast NIR technology over the past decade, developing cost-effective, high-performance and solution-processable NIR PDs that are fully compliant with regulations for consumer electronics are urgently needed. In this work, we develop PDs that fully address these requirements based on colloidal AgBiS$_2$ NCs. We hypothesized that the thickness of the photoactive layer is a critical parameter to achieve optimal performance. Thus, we finely tuned the NC film thickness through layer-by-layer (LBL) deposition from 20 nm to 115 nm, and determined the effect on the external quantum efficiency, dark current density and noise current. Having established devices giving high responsivities and specific detectivities, we sought to understand the influence of thickness on the response speed of the PDs under white light or NIR (940 nm wavelength) illumination. We rationalized the results obtained by analysing the drift lengths, transient responses, and transit times. From this, we identified ion migration as a hidden factor affecting the transient response of the PDs.

Furthermore, we quantified the activation energy barrier for ion migration, and how it influenced PD performance as a function of AgBiS$_2$ layer thickness. Finally, we demonstrated the practical applications of AgBiS$_2$ PDs by testing them as heartbeat sensors operating without encapsulation in ambient air.

## 2. Results and Discussion

### 2.1 Photodetector performance of AgBiS$_2$ PDs in the NIR region.

The AgBiS$_2$ NCs used in this work were synthesized through a modified hot-injection process[20] (see Experimental Section for details), and they had a mean size of 4±1 nm (Figure S1a, Supporting Information). These AgBiS$_2$ NCs have a rocksalt crystal structure, with both Ag$^+$ and Bi$^{3+}$ occupying the same crystallographic lattice site. Drop-cast AgBiS$_2$ NCs could retain their phase and visual appearance in ambient air with relative humidity ranging from 60-70% for over a month (Figure S2, Supporting Information), suggesting their high air stability.

Herein, AgBiS$_2$ PDs were fabricated as *n-i-p* photodiodes (**Figure 1**a), where ZnO and PTB7/MoO$_x$ were adopted as the electron (ETL) and hole transport layers (HTL), respectively, both of which show charge extraction level matching the band positions of AgBiS$_2$.[20,22] Ligand exchange treatment with tetramethylammonium iodide (TMAI) in methanol solution enabled the use of the layer-by-layer (LBL) method to fabricate PDs with controlled photoactive layer thickness (see details in Experimental Section). Our TMAI-treated AgBiS$_2$ films showed high absorption coefficients from the ultra-violet (UV) to NIR region, exceeding 10$^4$ cm$^{-1}$ for wavelengths below 1200 nm, and exceeding 10$^5$ cm$^{-1}$ for wavelengths below 775 nm (Figure S3, Supporting Information), making them suitable for NIR photodetection.

To achieve high performing PDs, the number of photoactive layers (*i.e.*, TMAI-treated AgBiS$_2$ NC films) was tuned from 2 to 9 to reduce the dark current density $J_D$. Herein, $J_D$ was only monitored at reverse bias smaller than 0.5 V (*i.e.*, $-V_A \leq 0.5$ V). Although a larger reverse bias (-1 ~ -10 V) is commonly used in commercial silicon PDs in order to extract charge-carriers out of their thick photoactive layers (hundreds of micrometres), it is not necessary for AgBiS$_2$ PDs because ultrathin photoactive layers (<120 nm) are sufficient for absorbing >65% of visible and near-infrared light (wavelengths <800 nm). Being able to operate these devices under smaller applied biases is advantageous for applications in wireless sensors as part of the Internet of Things, where reducing external power requirements is critical for these autonomous devices. The $J_D$–$V$ curves of AgBiS$_2$ PDs with 2–9 photoactive layers deposited (denoted as 2L – 9L) are shown in Figure S4a (Supporting Information), where we can see that there is an overall decrease in $J_D$ with an increase in the thickness of the photoactive layer. At the same time, AgBiS$_2$ PDs with different photoactive layers exhibited distinct external quantum efficiency (EQE) spectra (Figure S4b, Supporting Information), possibly caused by different optical interference effects. To compare the capability of these AgBiS$_2$ PDs in terms of NIR photodetection, the EQE values at 940 nm wavelength and the $J_D$ values biased at -0.5 V were simultaneously displayed in Figure S4c (Supporting Information), where we can clearly that the highest EQE and the lowest $J_D$ value, which are both desirable for an ideal NIR PD, were not obtained in the same photoactive layer. Therefore, to investigate how photoactive layer affects PD performance, AgBiS$_2$ PDs with 3, 5, and 9 photoactive layers (denoted as the 3L, 5L, and 9L AgBiS$_2$ PD) were characterized in detail, as will be discussed below. Herein, 5L and 9L AgBiS$_2$ PDs account for the devices showing the highest EQE values in the NIR region and the lowest $J_D$, respectively, while 3L AgBiS$_2$ PDs account for the thinnest devices among the three types as a reference.

The current density-voltage ($J$–$V$) curves of the AgBiS$_2$ PDs in the dark and under 1-sun illumination are displayed in Figure 1b, where we can see a $J_D$ value of 4.1×10$^{-5}$, 1.8×10$^{-5}$, and 0.37×10$^{-5}$ A cm$^{-2}$ at -0.5 V for the 3L, 5L, and 9L PD, respectively. Note that these $J_D$ values are comparable to those reported in some of the best NIR organic,[10,26,27] lead-halide perovskite[28,29] and PbS QD PDs[30,31] (see Table S1, Supporting Information). Figure 1c shows the responsivity ($R$) spectra of these devices. Since $R$ is proportional to the EQE value, it is not surprising that the 5L PD would display higher $R$ values in the NIR region (750 – 1080 nm) with a peak value of 0.28 A W$^{-1}$ at 770 nm.

Specific detectivity $D^*$ is another important metric for PDs because it determines the signal-to-noise ratio of PDs, which is related to $R$ by $D^* = \frac{\sqrt{A}R}{i_n}$ with $A$ and $i_n$ the photoactive area of the PD (4.5 mm$^2$), and the noise current spectral density, respectively. In the wider community, many researchers calculated $i_n$ assuming it is mainly comprized of the shot and thermal noise, as defined by Equation 1.[32,33]

$$i_n = \sqrt{2qi_D + \frac{4kT}{R_{\text{shunt}}}} \qquad (1)$$

In Equation 1, $q$ is the elementary charge, $i_D$ the static dark current, $k$ the Boltzmann constant, $T$ the temperature, and $R_{\text{shunt}}$ the shunt resistance. In practice, other sources of noise could contribute to $i_n$ as well. Therefore, using Equation 1 or even $\sqrt{2qi_D}$ only, which is also adopted in the literature,[30,34] would overestimate $D^*$ values, as depicted in Figure S6 (Supporting Information). To avoid such an overestimation, we determined the actual $i_n$ spectra of different devices by performing a Fast Fourier Transform (FFT) on the time evolution of the dark current (details in Experimental Section). As shown in Figure S5a (Supporting Information), thicker devices tended to show lower $i_n$ values, mainly due to their lower dark currents. The $i_n$ spectra for all of the devices were invariant with frequency, indicating that flicker noises, which are usually ascribed to the random trapping and de-trapping processes,[35] do not play a significant role in these devices. The calculated noise currents based on Equation 1 for all devices are also displayed in as the horizontal lines in Figure S5a (Supporting Information), and it can be seen that all of the measured noise currents were at least an order of magnitude higher than the values calculated from Equation 1. This result again verifies the importance of direct noise current measurements for precise $D^*$ determination. The tendency of the wider field working on solution-processed PDs to assume that the noise current only comes from shot noise, or a combination of shot and thermal noise, rather than directly measuring the noise current makes a comparison of $D^*$ between different reports difficult (see footnote in Table S1, Supporting Information).

The $D^*$ spectra of AgBiS$_2$ PDs were determined by their $R$ spectra and the directly-measured $i_n$ values, as shown in Figure 1d. Herein, the $i_n$ value was extracted from the high-frequency region above 1 kHz of the $i_n$ spectrum for each device biased at -0.5 V (Figure S5a, Supporting Information), which is estimated to be $8.9 \times 10^{-11}$, $1.3 \times 10^{-11}$, and $9.5 \times 10^{-12}$ A Hz$^{-1/2}$ for the 3L, 5L, and 9L PD, respectively. The 9L PD showed the highest $D^*$, with values over $10^9$ Jones in the region below 780 nm wavelength due to its lowest $i_n$ value, which is consistent with the lowest $J_D$ value in the 9L PD. However, the higher $R$ values in the NIR region of the 5L PD enabled it to exhibit the highest $D^*$ values (between $7 \times 10^8$ and $4 \times 10^9$ Jones) for wavelengths >780 nm. Additionally, the peak $D^*$ values of each PD (at 710 nm, 770 nm, and 620 nm wavelength for the 3L, 5L, 9L PD, respectively) calculated based on the $i_n$ values at different frequencies are also displayed in Figure S5b (Supporting Information), where we can see that the $D^*$ values are almost frequency-independent. This frequency independence further verifies the validity of the $D^*$ values here.

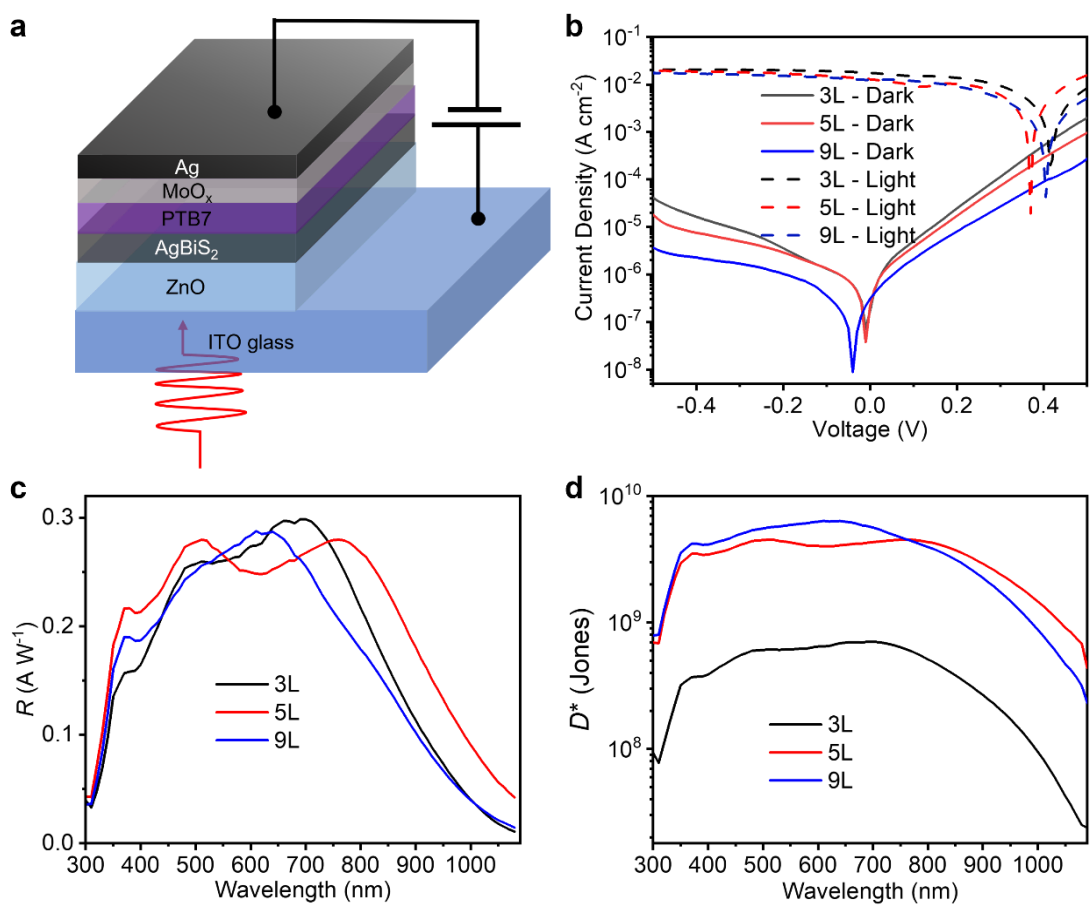

**Figure 1.** Performance of AgBiS$_2$ PDs. a) The *n-i-p* photodiode architecture of AgBiS$_2$ PDs. b) The dark current and photocurrent density under 1-sun illumination (AM1.5G spectrum) at different voltages ($J_D$-$V$ and $J_{ph}$-$V$ curves, represented by the solid and dash lines), c) responsivity ($R$) spectra, and d) specific detectivity ($D^*$) of AgBiS$_2$ PDs with 3 (3L), 5 (5L), and 9 (9L) photoactive layers. $D^*$ was determined using the directly-measured noise currents. If $D^*$ were instead calculated based on the dark current, as is common practice in the literature, they would be over an order of magnitude larger (Figure S6, Supporting Information).

Linear dynamic range (LDR) and film morphology were also characterized for our AgBiS$_2$ PDs. Under 940 nm light illumination, the 3L, 5L, and 9L PD showed LDR values of 35, 42, 52 dB, respectively (Figure S7, Supporting Information), which are comparable to organic PDs in the NIR region.[10,36] The slight decrease in LDR values for thinner devices mainly resulted from the increased $J_D$ values, which determined the lower limit of LDR. In addition, the film morphology of different layers of the TMAI-treated AgBiS$_2$ films deposited on top of the ZnO layer were characterized by atomic force microscopy (AFM). As displayed in Figure S8 (Supporting Information), all of the AgBiS$_2$ films showed a compact surface with a small root-mean-square (RMS) roughness (~1.15 nm, 0.13 nm, and 0.10 nm for the 3L, 5L, and 9L AgBiS$_2$ films, respectively). Although smaller roughness values might be observed in thicker films, we emphasize that the present roughness values achieved in all the samples have been comparable or even smaller than those reported in PbS QD (~0.5–5 nm),[37,38] organic (~0.5 nm–10 nm),[10,39] and lead-halide perovskite films (~10 nm–50 nm),[40,41] suggesting that the film morphology is not limiting the performance of these devices.

**2.2 Fast photo-response of AgBiS$_2$ PDs**

Having realized AgBiS$_2$ PDs with strong performance, we focused on understanding their photo-response speed in the NIR region and under broad-band white light illumination. Herein, cut-off frequenc $f_{-3\text{dB}}$ was characterized by tracking the damping of the generated photocurrents ($I_{\text{freq}}$) from the AgBiS$_2$ PDs under illumination with varying modulation frequencies, as given by Equation 2.

$$\text{Damping} = 20 \cdot \log_{10}\left(\frac{I_{\text{freq}}}{I_0}\right) \quad (2)$$

In Equation 2, $I_0$ is the photocurrent measured under continuous-wave (CW) light illumination. As the damping of the photocurrents drops to -3 dB, the associated frequency is $f_{-3\text{dB}}$. **Figure 2**a displays the damping measurements for AgBiS$_2$ PDs excited by a 940 nm LED driven by sinusoidal voltages at different frequencies. We can see that the $f_{-3dB}$ values were 244, 496, and 399 kHz for the 3L, 5L, and 9L PDs, respectively. The cut-off frequency of 496 kHz at 940 nm have exceeded the speed of organic PDs in the NIR region, and is also higher than almost all apart from one prior report of PbS QD PDs (Table S1, Supporting Information). We note that when PDs are illuminated by visible or UV light, they would usually display higher $f_{-3dB}$ values,[23,28,42] which might be due to higher diffusion currents as higher excess charge-carrier densities are created near the surface or increased filling of traps by the higher power densities achievable with shorter wavelength light sources. Indeed, this was also the case for the AgBiS$_2$ PDs in this work under white light (400–800 nm wavelength) illumination, which could reach $f_{-3dB}$ values over 1 MHz (1.20, 1.44, and 1.35 MHz for the 3L, 5L, and 9L devices, respectively). Not only does this cut-off frequency exceed the previously-reported value for AgBiS$_2$ PDs, it also exceeds nearly all of the reported cut-off frequencies for solution-processed NIR PDs (Table S1, Supporting Information).

**2.3 Rationalising the fast response of AgBiS$_2$ PDs**

It has been shown that the cut-off frequency $f_{-3dB}$ can be associated with charge-carrier mobility and film thickness by Equation 3.[26]

$$f_{-3dB} \propto \frac{V \mu_e \mu_h}{d^2 (\mu_e + \mu_h)} \tag{3}$$

In Equation 3, $V$ is the applied voltage across the two electrodes of the photodiode, $d$ the thickness of the photoactive layer, and $\mu_e$ and $\mu_h$ the electron and hole mobilities, respectively. In most solution-processed NIR PDs, $d$ is at least a few hundreds of nanometres in order to achieve sufficient light absorption,[7,15,34] and their $f_{-3dB}$ values would be thus limited by the long transit distances required. In addition, organic

PDs also suffer from low or unbalanced charge-carrier mobilities,[26,43] which could further lower their $f_{-3dB}$ values, since the charge-carrier transport will be limited by the smallest charge-carrier mobility. By contrast, the high absorption coefficients of AgBiS$_2$ NCs enable the fabrication of ultrathin devices here. As determined from atomic force microscopy (AFM), the AgBiS$_2$ film thicknesses in the 3L, 5L, and 9L PDs were only 19±3, 38±4, and 115±9 nm, respectively. Furthermore, calculations also predicted the comparable effective masses of electrons and holes (hole effective mass $m_h^*$ approximately only double the electron effective mass $m_e^*$)[44,45] in AgBiS$_2$, which could also be beneficial to achieving higher $f_{-3dB}$ value by having relatively balanced electron and hole transport.

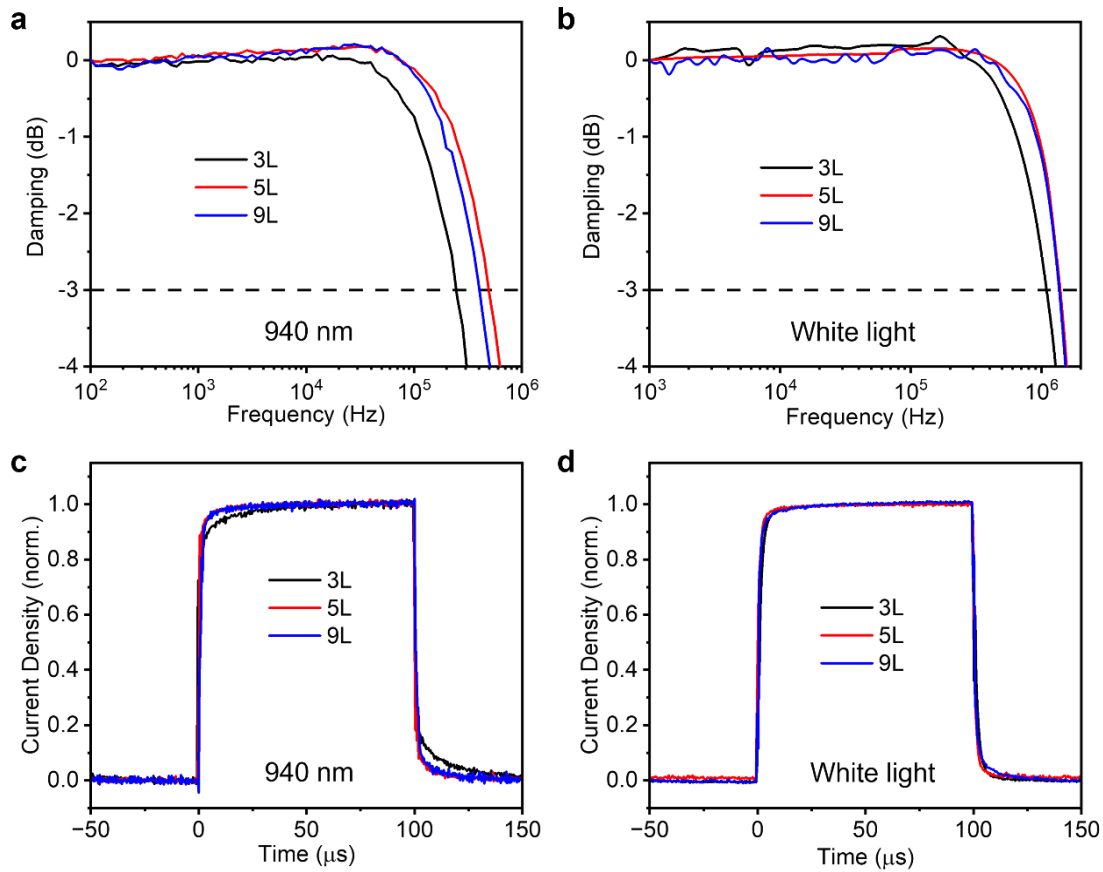

**Figure 2.** Fast response of AgBiS$_2$ PDs. Cut-off frequency measurements of the 3L, 5L, and 9L devices under the illumination from a) 940 nm as well as b) white light LEDs

driven by sinusoidal voltages at different frequencies. The normalized transient current densities of the 3L, 5L, and 9L devices under excitation from c) 940 nm and d) white light pulses emitted from LEDs. The pulse width was 100 μs for all the transient current density measurements. The devices were biased at 0 V in b) and at -0.5 V in other measurements.

More importantly, the ultrathin photoactive layers could facilitate efficient charge-carrier extraction out of AgBiS$_2$ PDs. An estimate of the drift length $L_{\text{drift}}$ can be obtained from Equation 4 from the diffusion length $L_{\text{diff}} = \sqrt{D\tau}$, where $D$ and $\tau$ are the diffusion constant and charge-carrier lifetimes, respectively. Since mobility $\mu$ can be associated with $D$ by Einstein relation, $L_{\text{drift}}$ can be expressed by Equation 4.

$$L_{\text{drift}} = v_{\text{drift}}\tau = \mu\left(\frac{V}{d}\right)\tau = \frac{eD}{kT}\left(\frac{V}{d}\right)\tau = \frac{e}{kT}\left(\frac{V}{d}\right)L_{\text{diff}}^2 \tag{4}$$

In Equation 4, $v_{\text{drift}}$ is the drift velocity. From time-resolved microwave conductivity (TRMC) measurements, the electron and hole diffusion lengths for TMAI-treated AgBiS$_2$ NC films have been reported to be approximately 60 and 150 nm, respectively.[46] Considering the AgBiS$_2$ PDs with different photoactive layers biased at -0.5 V, and assuming no screening of this field within the device, a $L_{\text{drift}}$ value of at least 3516, 1758, and 611 nm (for holes) could be estimated for the 3L, 5L, and 9L devices, respectively. We can see that all of the $L_{\text{drift}}$ values obtained from Equation 4 are substantially larger than the corresponding photoactive layer thicknesses in each device. Therefore, most photo-generated charge-carriers would be quickly swept across the photoactive layers under bias even in the thickest 9L AgBiS$_2$ PD. It has been reported in a PbS QD/metal Schottky diode, the response speed of the diode could be significantly improved when photo-generated charge-carriers were transported mainly by drift.[47] We note that a similar condition could be met in our AgBiS$_2$ PDs, where the thin photoactive layers were all included in the depletion regions so that drift transport would dominate and lead to a fast response.

The transient current densities of the AgBiS$_2$ PDs under pulsed illumination were also measured to study their rise/fall time $t_{\text{rise}}/t_{\text{fall}}$. The normalized transient current dynamics of the 3L, 5L, and 9L PD activated by light pulses emitted from a 940 nm and white light LED are displayed in Figure 2c and 2d, respectively. Here, rise/fall times refer to the time required for the transient currents to rise from 10% to 90% of their maximum values, and drop from 90% to 10% of the maximum values, respectively. The acquired $t_{\text{rise}}/t_{\text{fall}}$ values for the 3L, 5L, and 9L PD are listed in Table 1, where we can see response times shorter than 10 μs in all the devices under either 940 nm or white light illumination. It can be seen from Table S1 (Supporting Information) that the acquired rise/fall times in AgBiS$_2$ PDs are shorter than most solution-processed PDs in the NIR region and even comparable to some lead-halide perovskite-based PDs in the blue light region.[42,48]

In principle, the charge-carrier transit time $t_{\text{tr}}$ through a PD can be estimated from Equation 5.[49]

$$t_{\text{tr}} = \frac{A_0}{\sqrt{(2\pi f_{-3\text{dB}})^2 - \left(\frac{1}{R_t C}\right)^2}} \tag{5}$$

In Equation 5, $A_0$ is a constant typically equal to ~3.5, and $R_t C$ refers to the effective resistance-capacitance product of the overall experimental system. The series resistance of the AgBiS$_2$ PDs could be estimated from the reciprocal of the slope of the photocurrent density-voltage curves ($J_{\text{ph}}$-$V$ curves) at open-circuit voltage ($V_{\text{OC}}$) point (dash lines in Figure 1b), and the effective resistance of the experimental system was then estimated by the summation of the series resistance of each device and the connected resistance within the oscilloscope (50 Ω). The $C$ values were directly measured from the high-frequency range of the capacitance-frequency plots (Figure S9, Supporting Information). The $R_tC$ values for different AgBiS$_2$ PDs are listed in Table 1, from which we can calculate the transit time $t_{tr}$.

**Table 1.** The rise/fall time $t_{rise}/t_{fall}$, cut-off frequency $f_{-3dB}$, $1/R_tC$ values with $R_tC$ the effective resistance-capacitance product, and estimated transit time $t_{tr}$ of AgBiS$_2$ PDs.

| Device | $t_{rise}$; $t_{fall}$ (μs) | $f_{-3dB}$ (Hz) | $1/R_tC$ (Ω$^{-1}$ F$^{-1}$) | $t_{tr}$ (μs) |
|---|---|---|---|---|
| 3L | 5.2/8.3 (940 nm); 3.7/3.6 (white light) | 244 k (940 nm); 1.20 M (white light) | $8.50 \times 10^4$ | 2.3 (940 nm); 0.5 (white light) |
| 5L | 1.9/1.9 (940 nm); 2.6/2.5 (white light) | 496 k (940 nm); 1.44 M (white light) | $2.71 \times 10^4$ | 1.1 (940 nm); 0.4 (white light) |
| 9L | 2.1/2.5 (940 nm); 3.0/3.7 (white light) | 399 k (940 nm); 1.35 M (white light) | $2.05 \times 10^4$ | 1.4 (940 nm); 0.4 (white light) |

As can be seen from Table 1, the calculated $t_{tr}$ values were close to the $t_{rise}/t_{fall}$ values under 940 nm light illumination, suggesting that the device response times were mainly determined by the charge-carrier transit times through the devices. On the other hand, under white light illumination, the $t_{rise}/t_{fall}$ values tended to be slightly longer than the calculated $t_{tr}$ values for each device. Indeed, smaller $t_{rise}/t_{fall}$ values were initially expected to be reached under white light illumination (larger $f_{-3dB}$ values) and also in thinner devices (larger drift velocities for charge-carriers), both of which were not the case, as can be seen from Table 1. These unusual phenomena imply the influence of a hidden factor – ion migration, might occur during the characterization. Here, we propose that ion migration is facilitated by a stronger electric field in thinner devices, or by the higher light intensity from the white light source, which could potentially induce a large charge-carrier gradient and hence drive ions migration more easily.[50,51] It is therefore important to understand the role of ion migration in $AgBiS_2$, which is examined in the next section.

**2.4 Ion migration in $AgBiS_2$**

Ion migration has been observed in several perovskite and PbS QD-based devices,[52–54] where mobile cations or halogen ligands could easily move when driven by the external electric field. In order to verify whether ion migration could also occur in $AgBiS_2$, temperature-dependent transient dark current measurements (see details in Experimental Section) were performed here. At higher temperatures, more ions will be thermally activated[55] and accumulated near the electrodes, which would gradually screen the electric field between the two electrodes. As a result, the transient dark currents excited by voltage pulses would decay faster when the external field was screened more by the built-in field from the accumulated ions.

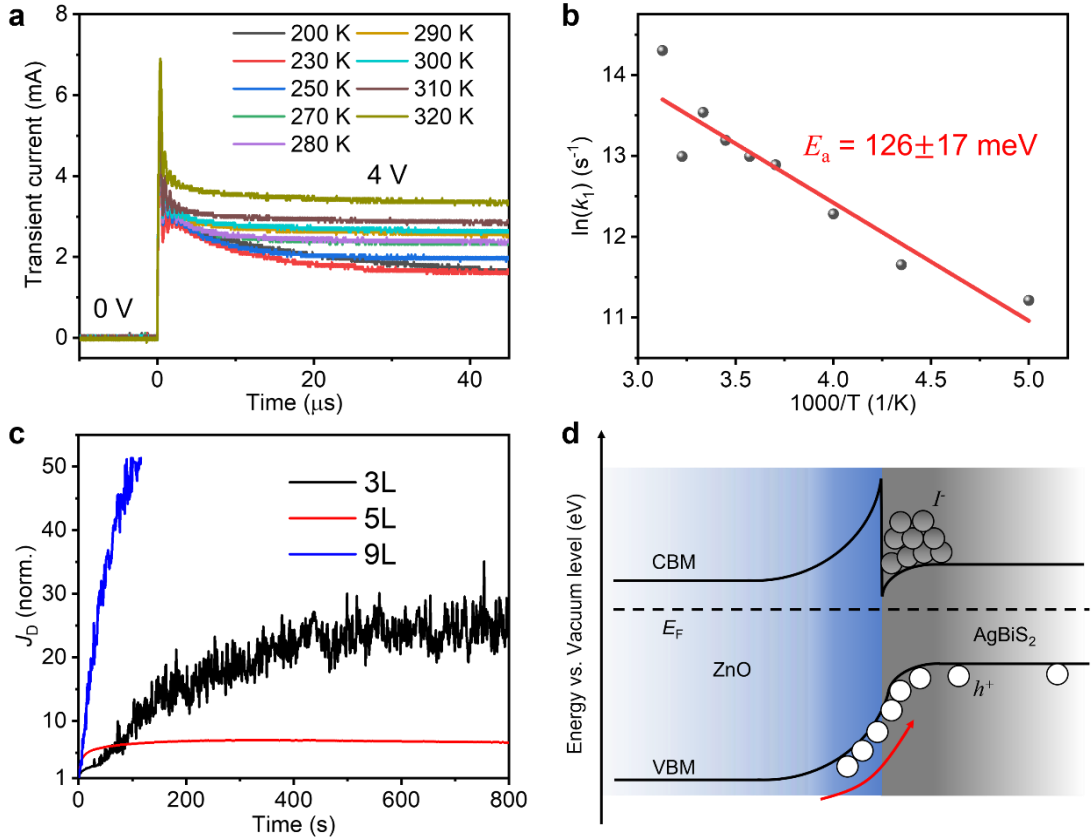

**Figure 3.** Understanding ion migration in AgBiS$_2$. a) Temperature-dependent transient dark currents of the electron-only 3L AgBiS$_2$ device (ITO/ZnO/AgBiS$_2$/PCBM/Ag) activated by voltage pulses (pulse height: 4 V, pulse width: 50 μs) from 200 K to 320 K, and b) the corresponding Arrhenius plot, along with the extracted activation energy barrier, $E_a$. $k_1$, and $T$ refer to the decay rate of the transient currents and temperature, respectively. c) Time evolution of the normalized dark current density $J_D$ of the 3L, 5L, and 9L devices biased at -0.5 V. All the time evolution profiles were normalized to the initial $J_D$ values at time = 0 s. d) Proposed band alignment and bending at the ZnO/AgBiS$_2$ heterojunction. CBM, VBM, $E_F$, $I^-$ and $h^+$ refer to the conduction band minimum, valence band maximum, Fermi level, accumulated ions, and injected holes, respectively.

To simplify the analyses of ion migration in AgBiS$_2$, electron-only devices (architecture:

ITO/ZnO/AgBiS$_2$/PCBM/Ag) with 3 photoactive layers (3L) were used here. An electron-only device was used because of the *n*-type nature of TMAI-treated AgBiS$_2$ films, as verified from Kelvin probe (KP) and photoelectron yield spectroscopy (PYS) measurements (Figure S10, Supporting Information). **Figure 3**a shows the dark current transients of the electron-only AgBiS$_2$ device excited by high-voltage (4 V) pulses from 200 K to 320 K, where we could indeed see the transients decaying faster at higher rates $k_1$ (Table S2, Supporting Information) as temperature was increased. The corresponding Arrhenius plot (analysis details in Note S1, Supporting Information) of AgBiS$_2$ is then displayed in Figure 3b, and an activation energy barrier ($E_a$) of approximately 126±17 meV was extracted. This activation energy barrier is comparable to or smaller than those reported in lead-halide perovskites (typically in the range of 100–300 meV, with some reports of up to 600 meV[53,54,56,57]), where ion migration has been proven to strongly influence their device performance. We thus conclude here that ions could easily migrate in AgBiS$_2$ PDs, and possibly impact the photo-responses more significantly under a stronger electric field (*e.g.*, in the 3L device) or higher light intensity[58] (*e.g.*, under white light illumination).

To examine the possible impact of ion migration on the dark current density $J_D$ of AgBiS$_2$ PDs over a longer time scale, we tracked the dark current density $J_D$ of the 3L, 5L, and 9L AgBiS$_2$ PDs under continuous bias for a few minutes. As shown in Figure 3c, we noticed that the initial $J_D$ value of the 3L and 5L devices were enhanced by a factor of 24 and 7, respectively, after 13 min, while the 9L device $J_D$ was increased by a factor of 51 after only 2 min. Such an enhancement in $J_D$ values suggests ion migration to occur in all the AgBiS$_2$ PDs biased for a long time scale. Potentially, the accumulated ions at the AgBiS$_2$/transport layer interfaces could induce the charge-injection from the electrodes, and therefore enhance the dark currents. Moreover, KP and PYS measurements also indicated that our synthesized ZnO films had a lower electron affinity than the TMAI-treated AgBiS$_2$ film (Figure S10, Supporting Information; *i.e.*, negative conduction band offset). This could lead to a contact barrier (type-II heterojunction) at the ZnO/AgBiS$_2$ interface in these devices. Although some charge-carriers may still tunnel through this barrier *via* the tail of states extending from the ZnO band edges[59] (which are also shown to be present from the broadband photoluminescence emission of the ZnO films; Figure S11, Supporting Information), the negative conduction band offset could make ions accumulate more easily and hence encourage the hole injection from ZnO under reverse bias (Figure 3d). Such a $J_D$ enhancement over long-term biasing was also observed in tetrabutylammonium iodide (TBAI)-treated PbS QDs,[60] and iodine anions have been claimed to be the main source for ion migration owing to their weaker binding energy with PbS QDs. Considering the similar crystal structure and iodide ligand used for AgBiS$_2$ films, iodine anions may also be the main source of migrating ions in AgBiS$_2$ PDs as well.

Given that ions are driven by the application of an external field, thinner devices were expected to exhibit more significant ion migration, with a larger increase in $J_D$ over time under continuous reverse bias. Therefore, it is not surprising to see a larger $J_D$ enhancement and slower photo-response in the 3L device (Figure 3). Interestingly, although the 9L device had the smallest $J_D$ at -0.5 V (Figure 1b), its $J_D$ enhancement is still large compared to other devices, suggesting that the applied field might not be the only factor facilitating ion migration. As a result, the smallest $J_D$ enhancement factor was actually seen from the 5L device, which could have benefited from 1) not having as high an electric field as the 3L device, and 2) not having a large concentration of defects introduced during the ligand-exchange process, as explained in Figure S12 (Supporting Information). The small $J_D$ enhancement in the 5L device indicates the minor effect of ion migration in this device, which also contributes to its fastest photo-response (Figure 2).

Herein, we have shown that although ion migration could be prevalent in $AgBiS_2$, its effect can be effectively suppressed by fine-tuning the photoactive layer thickness in $AgBiS_2$ PDs. We note that the initial $J_D$ value in the 5L PD was only enhanced by less than one order of magnitude (7 times) under continuous -0.5 V bias, which had minor impact on both noise current and specific detectivity.

### 2.5 Real-time heartbeat monitoring using $AgBiS_2$ PDs

To demonstrate the practical application of $AgBiS_2$ PDs, we performed real-time photoplethysmogram measurements using the 5L device. As shown in **Figure 4**a, a finger of a volunteer was placed between a 940 nm LED and a 5L $AgBiS_2$ PD. The pulsating bloodstream through the capillaries changed the transmission of 940 nm NIR light from an LED. In the NIR region, not only could light penetrate living tissue efficiently, but also the difference in absorbance between oxygenated and deoxygenated haemoglobin will be maximized. As a result, the heartbeat rate of the volunteer could be precisely recorded as the electric signals by the 5L PD, as shown in Figure 4b. An even clearer figure could be made when we differentiated the current signals with time (Figure 4c), where we could easily estimate a heartbeat rate of around 60 beats per minute. Finally, it is worth mentioning that unlike lead-halide perovskite PDs, where $J_D$ would be enhanced by two times after being storing in air for 10 days,[61] the $J_D$ value of the unencapsulated $AgBiS_2$ PD was not increased when stored in an ambient environment (relative humidity: 60–70%) over 34 days (Figure S13, Supporting Information). This result also suggests that $AgBiS_2$ PDs are air-stable, which is consistent with the high stability of the $AgBiS_2$ solar cells reported previously.[20,21]

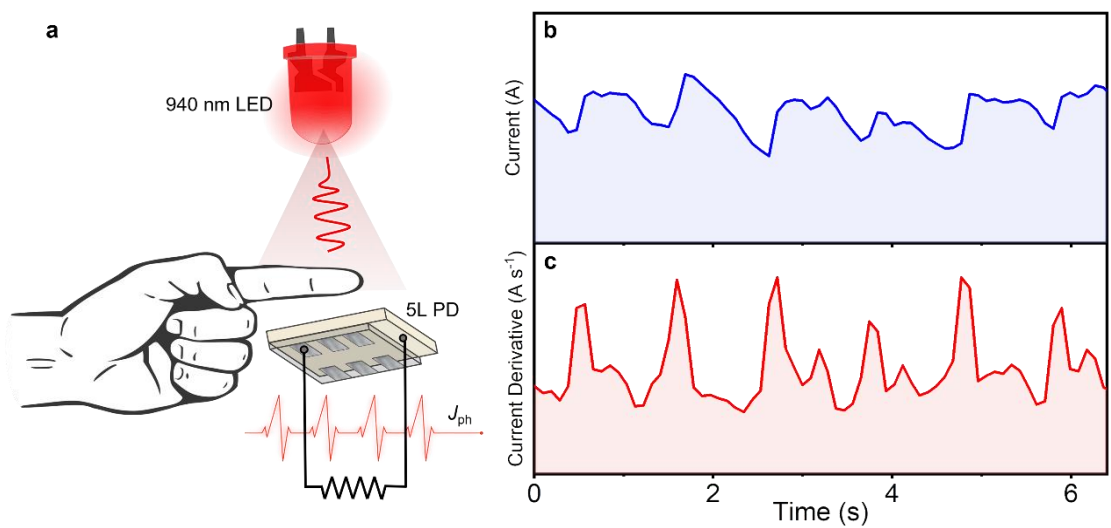

**Figure 4.** AgBiS$_2$ PDs as NIR heartbeat sensors. a) Schematic illustration of the real-time photoplethysmogram setup, and b) measurements of the heart beat of a finger under 940 nm light illumination, along with c) its time derivative. The signals were recorded using the 5L AgBiS$_2$ PD.

## 3. Conclusion

In this work, AgBiS$_2$ NCs have been shown to overcome the shortcomings of current solution-processed absorbers for NIR photodetection. A fast response, high stability in air, along with a composition of nontoxic, widely-available elements were simultaneously realized in AgBiS$_2$ PDs. Low $J_D$ values (< 10$^{-5}$ A cm$^{-2}$) as well as high $D^*$ values in the NIR region (>10$^9$ Jones) could be achieved in AgBiS$_2$ PDs with a finely-tuned photoactive layer thickness, and the strong absorption of AgBiS$_2$ films also enabled the fabrication of ultrathin devices, which further led to high $f_{\text{-3dB}}$ values approaching 500 kHz in the NIR region and over 1 MHz under white light illumination. Such a fast photo-response is essential for various high-throughput applications, such as optical communications, but are challenging to achieve in most solution-processed

PDs, mainly due to their bulkiness and poor charge-carrier transport. In addition, ion migration has been shown to occur in AgBiS$_2$ PDs with a small activation energy barrier of 126 meV, which could potentially delay the photo-response or increase dark currents over a large time scale. However, the impacts of ion migration could be effectively mitigated by using an intermediate photoactive layer thickness. These remarkable features make AgBiS$_2$ PDs particularly suitable for practical applications, and they have been demonstrated as efficient heartbeat sensors here. Our work therefore pushes forward AgBiS$_2$ as a new opportunity for fast and solution-processable next-generation NIR PDs.

**Acknowledgements**

Y.-T.H. would like to thank the Ministry of Education, Taiwan and Downing College Cambridge for funding. N.G. and F.F. would like to acknowledge support from European Commission Research Executive Agency (Grant Agreement Number: 859752 HEL4CHIR-OLED H2020-MSCA-ITN-2019). L.D. thanks the Cambridge Trusts and the China Scholarship Council for funding. Y.-T.H and R.L.Z.H. acknowledge support from the Silverman Research Fellowship, Downing College Cambridge. R.L.Z.H. would like to thank EPSRC for financial support (grant no. EP/V014498/2), as well as the Royal Academy of Engineering through the Research Fellowships scheme (no. RF \201718\17101).

**Author Contributions**

Y.-T.H., N.G. and R.L.Z.H. conceived of this work, and Y.-T.H. and D.N. contributed equally to this work. Y.-T.H. developed and optimized the synthesis of AgBiS$_2$ NCs, fabricated AgBiS$_2$ devices, and performed XRD, AFM, stability tests as well as temperature-dependent transient current measurements under the supervision of A.R. and R.L.Z.H. D.N. and F.F. performed PD characterisation under the supervision of N.G. Y.Z. assisted the temperature-dependent transient current measurements under the supervision of H.S. M.R. performed the KP and PYS measurements. L.D. took the TEM images under the supervision of S.D.S. Z.A.-G. performed the PDS measurements. All authors discussed the results and wrote the paper together.

**Competing Interests**

The authors declare no financial or non-financial competing interests.

Fast near-infrared (NIR) photodetectors are vital for many applications, including optical communications and data transfer. Herein, we develop solution-processed AgBiS$_2$ photodetectors with a cut-off frequency approaching 500 kHz under NIR (940 nm) illumination, and >1 MHz under visible light. This comes about because of the high absorption strength of this material, enabling the fabrication of ultrathin photodetectors with fast charge-extraction, and we demonstrate their utility as heart-beat sensors in air.

Y.-T. Huang, D. Nodari, F. Furlan, Y. Zhang, M. Rusu, L. Dai, Z. Andaji-Garmaroudi, S. D. Stranks, H. Sirringhaus, A. Rao, N. Gasparini,* and R. L. Z. Hoye*

**Fast near-infrared photodetectors based on nontoxic and solution-processable AgBiS$_2$**

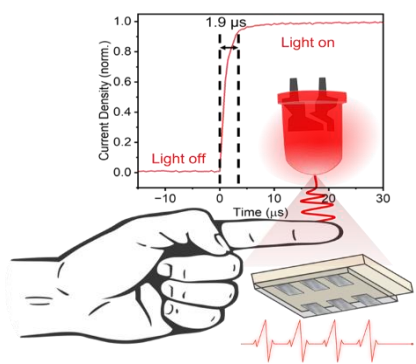